\documentclass[twocolumn,prl,superscriptaddress,showpacs]{revtex4}
\usepackage{amsmath,amssymb}
\usepackage{graphicx}
\usepackage{float}
\usepackage{subfigure}
\usepackage{verbatim}
\usepackage{amsfonts}
\usepackage{dcolumn}
\usepackage{bm}
\usepackage{color}
\usepackage[colorlinks,citecolor=blue]{hyperref}

\begin{document}

\hyphenpenalty=5000
\tolerance=1000

\title{A Study of Highly Frustrated Spin Systems with mixed PEPS in Infinite Honeycomb Lattice}

\author{Huan He}
\affiliation{Key Laboratory of Quantum Information, CAS, University of Science and Technology of China}
\author{Zhen Wang}
\affiliation{Key Laboratory of Quantum Information, CAS, University of Science and Technology of China}
\author{Chuanfeng Li}
\affiliation{Key Laboratory of Quantum Information, CAS, University of Science and Technology of China}
\author{YongJian Han
\footnote{Email address: smhan@ustc.edu.cn}}
\affiliation{Key Laboratory of Quantum Information, CAS, University of Science and Technology of China}
\author{Guangcan Guo}
\affiliation{Key Laboratory of Quantum Information, CAS, University of Science and Technology of China}

\begin{abstract}
Highly frustrated spin systems represent a central and challenging problem in condensed mater physics. To this problem, we introduce an algorithm based on mixed projected entangled pair states (\emph{m}-PEPS), which is a novel type of tensor network. We use the famous Kitaev model on an infinite honeycomb lattice, which can be solved exactly, as a benchmark. With very limited parameters and finite scaling, our calculation results are in good agreement with the exact results, indicating the efficiency of our algorithm. After presenting the benchmark, we investigate the Kitaev-Heisenberg model, which was proposed to describe the effective magnetic momentum interaction in iridate Na$_2$IrO$_3$ which may be used to realize the spin liquid phase. However, our calculations suggest that the gapless spin liquid phase is not robust at the thermodynamic limit, and thus this phenomenon is very difficult to observe.
\end{abstract}


\maketitle

Frustrated spin systems play an essential role in two-dimensional many-body physics and have attracted the attention of condensed matter physicists for decades\cite{balents}. In particular, some exotic ground states, such as the quantum spin liquid state (which has recently been strongly supported experimentally in the Kagome lattice by Y. Lee\cite{experi} \emph{et al}), may play a key role in explaining high-$T_c$ superconductivity, as proposed by P. W. Anderson\cite{and}. The Kitaev model on a honeycomb lattice\cite{kitaev}, which can be solved exactly and supports a spin liquid phase, is an important frustrated model in two-dimensional physics. In addition, this proposed model was proposed that it may be realized in a cold atom system\cite{luming}, or in the iridate Na$_2$IrO$_3$ \cite{KHmodel}. The exact solution shows that the spin liquid phase is gapless, and a gap opens when a magnetic field, which breaks the time-reversal symmetry, is applied to this model as a perturbation. Unfortunately, general two-dimensional frustrated spin models - even slightly modified Kitaev models such as the Kitaev model perturbed with a magnetic field - cannot be solved exactly.

Due to the lack of analytical methods for general many-body systems, numerical analysis remains the main method for understanding such physics. However, traditional algorithms, such as the quantum Monte Carlo (QMC) method and density matrix renormalization group (DMRG) method, have failed to simulate frustrated systems in two dimensions: QMC suffers from the notorious `sign problem' \cite{sign} while DMRG is limited to one dimension and generally lose its power in higher dimensional systems\cite{dmrg1,dmrg2}. Fortunately, the recently developed tensor network (TN) algorithms, such as, the algorithm based on projected entangled pair states (PEPS)\cite{PEPS1,PEPS2,PEPS3} which is a natural generalization of the DMRG algorithm to higher dimensions, has shown great potential for addressing this problem.  The algorithm based on infinite PEPS has been used to obtain a reasonable phase diagram of the frustrated antiferromagnetic $J_1-J_2$ Heisenberg model on a checkerboard lattice\cite{chan}. More recently, a spin liquid phase was claimed near the maximally frustrated region ($J_2\sim 0.5J_1$) for the spin $1/2$  $J_1-J_2$ antiferromagnetic Heisenberg model on a square lattice for the TN algorithm\cite{wangling,hongcheng}. These results show that the TN algorithms are powerful tools for exploring the frustrated systems and for evaluating some new physics beyond the traditional methods.

Despite the success of the TN algorithm, the study of highly frustrated large systems, such as those represented by the Kitaev model and related models, remains a challenge. The general TN algorithm, such as the algorithm based on PEPS, cannot give the appropriate results for the Kitaev model, because the limited ability of the calculation constrains the bond dimension $D$ to a relatively small number (typically $< 10$). Thus new methods must be developed to highly frustrated systems with the current calculation abilities. In this letter, we introduce an algorithm based on infinite mPEPS, which is a mixture of the bosonic PEPS (bPEPS)\cite{PEPS2} and the fermionic PEPS (fPEPS)\cite{fPEPS1,fPEPS2}. We will demostrate that this novel method can be used to efficiently study highly frustrated systems with a relatively small parameter $D$. We use the Kitaev model as a benchmark and obtain satisfactory results, for both the energy, and the correlations of this model. After the benchmark calculation, we apply this method to the Kitaev-Heisenberg model, which is proposed to describe the effective magnetic momentum interaction model of the iridate Na$_2$IrO$_3$ \cite{KHmodel,comment1}, to obtain a phase diagram. Our calculations indicate that the gapless spin liquid region is very narrow in the thermodynamic limit and that would thus be very difficult to observe in the iridate Na$_2$IrO$_3$, even with the Kitaev -Heisenberg model.


\textbf{Mixed Projected Entangled-Pair States} The TN algorithm is a variational method based on TN states with special structures. A general TN state is defined in two steps: first, we describe a configuration of entangled states between virtual particles (we call each entangled state a bond, connected by a line, as shown in Fig.1), which determines the structure and the upper bound of the entanglement of a TN state; second, we define projectors to project the virtual-particle space into real physical space. The different TN algorithms are distinguished by the structure of the TN state, such as PEPS, or the string bond state. The variational parameter space of TN algorithms is determined by the parameters of the projectors, the number of which is dependent on a polynomial of the dimension of the bond $D$, the dimension of the physical space $d$ and the number of projectors (as determined by the structure of the lattice and the symmetry of the state). For the special TN state, PEPS, the configuration of the bonds is completely determined by the lattice where each edge of the lattice corresponds to a bond, and each site has an on-site projector. Without a loss of generality, we use the honeycomb lattice as an example. In general, the entangled pair can be written as: $\mbox{EPR}_{e}=\sum_{i}{|i\rangle_{e_1}|i\rangle_{e_2}}$, where $|i\rangle$ is the state of virtual particle with dimension $D$ and $e$ denotes an edge connecting sites $e_1$ and $e_2$ in the lattice. Algebraically, the state $|i\rangle$ can be represented by local creation operators from the vacuum through the second quantization method. In the original PEPS case, the entangled pair state can be represented as bosonic operator $a^{\dag}$ as: $\mbox{EPR}_{e}=\sum_{i}{a^{\dag i}_{e_1}a^{\dag i}_{e_2}|Vac\rangle}$.

Similarly, the projector of each site on a honeycomb lattice can also be represented as bosonic operators as: $P^{[l_1l_2]}=\sum_{p,i,j,k}T^{[l_1l_2]}_{pijk}b^{\dag p}a_x^ia_y^ja_z^k|vac\rangle\langle Vac|$,
where $[l_1l_2]$ is the coordinate of the site, $b^\dag$ is the bosonic creation operator of the physical particle, and $a_x$,$a_y$,$a_z$ are the bosonic annihilation operators of the virtual particles connected to edges $x$,$y$ and $z$, respectively (see Fig.1). In this case $|vac\rangle$ is the vacuum of the physical space, and $|Vac\rangle$ is the vacuum of the virtual space. Using the representation of the entangled pairs and projectors, the PEPS state can be represented as: $|\psi\rangle=\prod_{l_1,l_2}P^{[l_1l_2]}\prod_e \mbox{EPR}_e$, where all of the virtual particles are projected as physical particles.

In general, the virtual particles in an \mbox{EPR} pair are not necessarily bosons. In solving the many-body physics of fermionic systems, it is natural to extend the bPEPS formulism to the fPEPS formulism. This can be completed by replacing the bosonic operators in the entangled pairs and projectors, as shown in Eq.(1) with fermionic operators. In the fPEPS formulism, the parity of the projectors at different sites of the lattice, which determine the exchange character of the projectors, should be the same even. This requirement guarantees that the fPEPS is well defined, that is, the order of the projectors is not important for the state beyond the global phase.

Here we focus on a frustrated spin system, which exhibits bosonic statistics in physical space. Thus, we propose an algorithm based on the mixed PEPS (mPEPS), which introduces fermionic statistics to the virtual-particle space (same as fPEPS) and bosonic statistics in the physical-particle space (different from fPEPS). Therefore, the projectors of the mPEPS should be expressed in the second quantized form as: $\sum_{p,i,j,k}T_{pijk}b^{\dag p}a_x^ia_y^ja_z^k|vac\rangle\langle Vac|$, where $b^\dag$ satisfies: $[b_i,b^\dag_j]=\delta_{ij}$ and $a_x$,$a_y$,$a_z$ satisfy: $\{a_i,a^\dag_j\}=\delta_{ij}$. To validate this definition, the parity of the fermionic operators in the projector should also be same for every site (for example, it could be even). Although there is some overlap, the state spaces of the PEPS, fPEPS and the mPEPS are expected to be different, even for the same parameters.

After the defining mPEPS, the ground state of a system in variational space can be found by the imaginary evolution method which is performed in the same manner as in the PEPS method; in particularly, we use the simple update method to approximate the environment in the same manner as PEPS in two dimensions\cite{xiangtao}. When the ground state is stable under the imaginary evolution, the physical quantities of the system can be calculated through a complex contraction process (for more details, see the Supplementary Information). In the contraction, the bosonic and fermionic operators must be carefully handled(see in the Supplementary Information).

\begin{figure}[H]
 \centering
 \includegraphics[width=0.4\textwidth,scale=0.8]{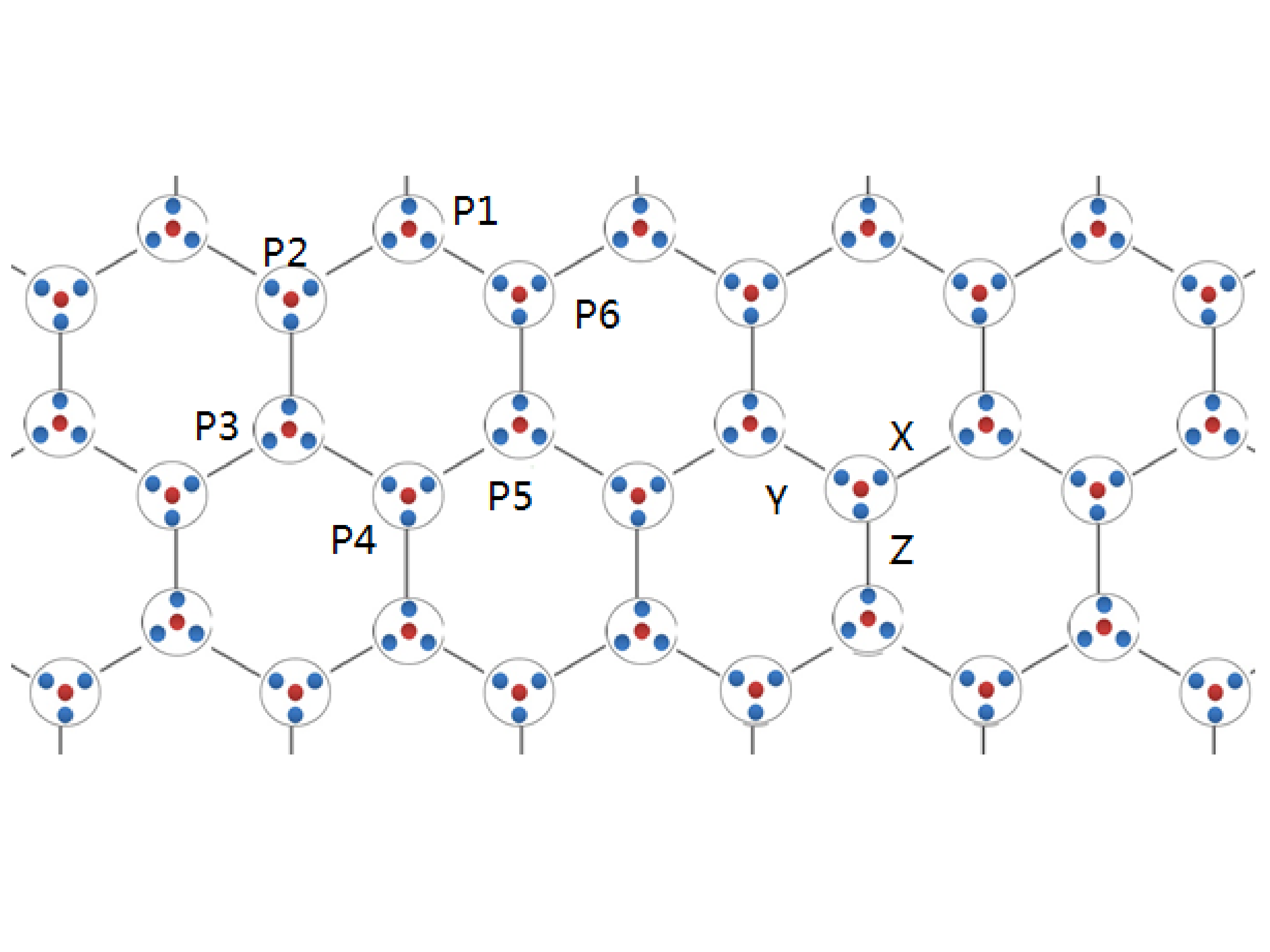}\\
 \caption{(Color on line) Representation of the mPEPS on a honeycomb lattice: each line denotes an entangled pair state; the blue balls denote virtual fermions; and the red balls denote bosons; the circle around each site represents an on-site projector $P$. Different link directions are denoted by $x$,$y$ and $z$, which are used in the Kitaev model. In the following calculations, we use an infinite lattice with translational symmetry and six different projectors, $P_1,P_2,\cdots,P_6$}
 \label{fig.1}
\end{figure}

\begin{figure}[H]
 \centering
 \includegraphics[width=0.5\textwidth]{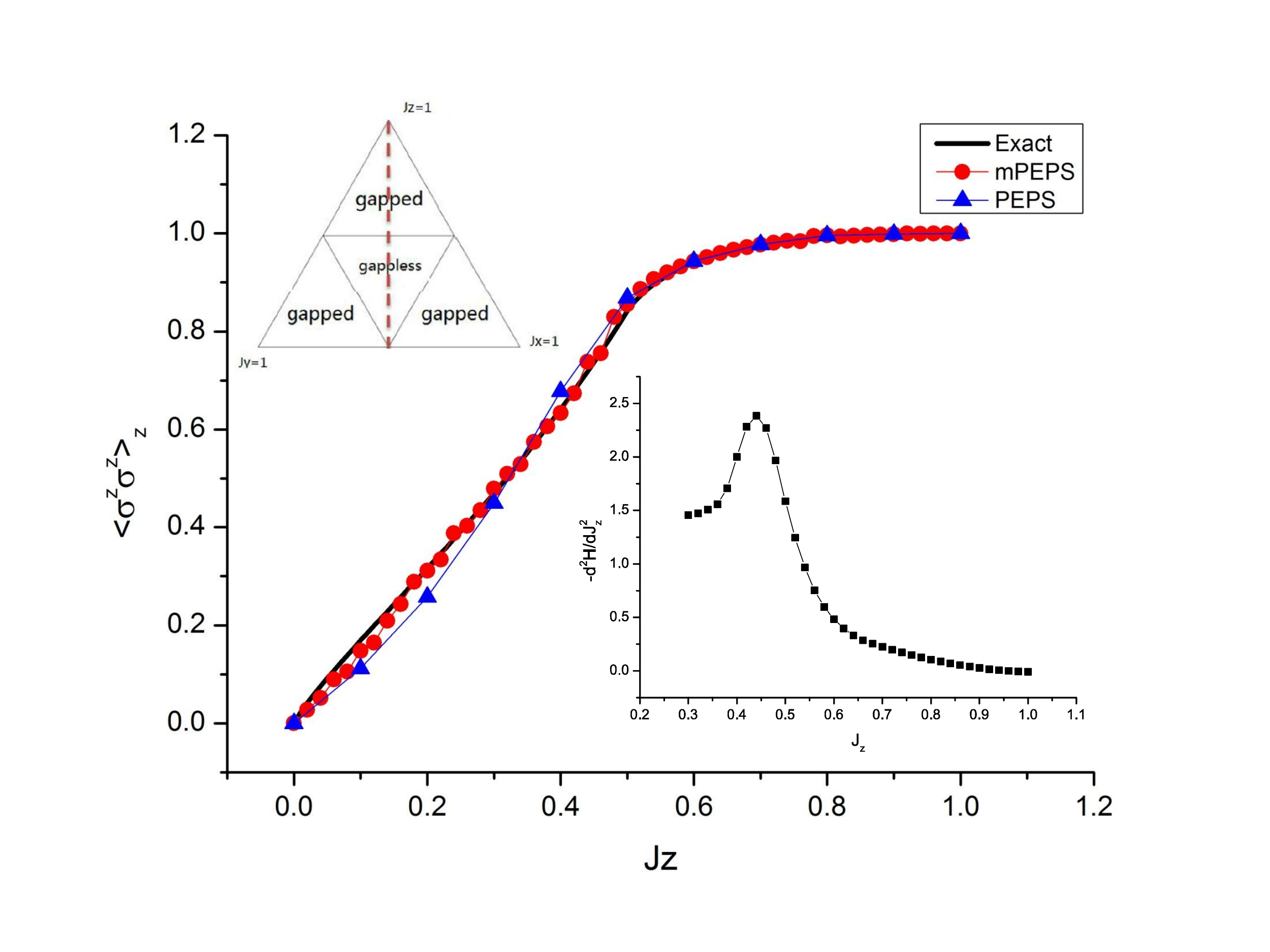}
 \caption{(Color on line) Nearest Neighbor(NN) correlation $\langle \sigma^z\sigma^z\rangle_z$ of the Kitaev model along the line $J_x=J_y$, which is indicated in the left upper inset by a dashed line. The red dots represent the results obtained by the mPEPS method with finite scaling of $D$; the black line shows the exact results of $\langle\sigma^z\sigma^z\rangle$ which is given as the formula $\langle \sigma^z_i\sigma^z_j\rangle=\frac{\sqrt 3}{16\pi^2}\int_{BZ}cos\theta(k_1,k_2)dk_1 dk_2$\cite{baskaran}; the blue triangles display the results obtained by the PEPS method using the same parameters of mPEPS. The right lower inset shows the $\langle\frac{\partial^2 H}{\partial J^2}\rangle$, which displays the transition point near 0.46.
 }
 \label{fig.2}
\end{figure}

\textbf{Kitaev Model as the Benchmark} Here, we use the Kitaev model, which is a highly frustrated spin model in two dimensions, as a benchmark of the mPEPS algorithm. The Kitaev model is defined on an infinite honeycomb lattice\cite{kitaev}, and its interaction along different directions (see Fig.1) vary strongly as: $H=\sum_{\gamma-link}J_{\gamma}\sigma^{\gamma}_i\sigma^{\gamma}_j$. This model can be exactly solved with two resultant phases: the gapped phase and the gapless spin liquid. There are two types of quasi-particles in the gapped phase: fermions and vortices. The vortices associated with a $Z_2$ gauge field and a factor $-1$ will appear when the fermion moves around the vortex. A gap opens when a magnetic field is applied to the gapless spin liquid phase. The non-abelian anyon, which may play an important role in topological quantum computation, will also appear in this phase. In our calculation, we normally set the parameters of the model to satisfy $J_x+J_y+J_z=1$ and $J_x=J_y$, which gives a line connecting a vertex to the middle point of the corresponding edge (see the red dashed line in the inset of Fig.2). After conducting the stability tests of the algorithm with respect to the parameters (particularly, the truncation parameter $D_{cut}$ in the contraction process, see details in the Supplementary Information), we compare the energy of the system along the chosen line with the exact solution. Result show good agreement (see the Supplementary Information). In addition, the nearest neighbor(NN) correlation $\langle\sigma^z\sigma^z\rangle$ which corresponds to $\langle\frac{\partial H}{\partial Jz}\rangle$ along the line is also shown in the Fig.\ref{fig.2}. The black line shows the exact value of this correlation as derived in\cite{baskaran}, which is in good agreement with our calculation (red dots). The inset shows the $\langle\frac{\partial^2 H}{\partial Jz^2}\rangle$, which exhibits a sharp transition point near 0.46 and indicates a phase transition in this model, although it slightly deviates from the exact result of 0.5.

We must emphasize that the virtual dimension of the system $D$ is rather limited by our calculation ability. We can only calculate results for the dimension up to 8. And we conduct the finite scaling for this limited dimension in this model, and the points shown in Fig.\ref{fig.2} were obtained with those finite scaling. To compare our findings with the traditional PEPS case, we also present the results from the PEPS with the same parameters and finite scaling employed in Fig.\ref{fig.2} (blue triangles). It is clear that the present method improves the results obtain with limited calculation ability. More importantly, the PEPS method does not give an indication of the phase transition, while our mPEPS technique can. However, in simpler cases, such as two-dimensional Ising model with a transverse magnetic field on the honeycomb lattice, the mPEPS and PEPS methods give nearly the equivalent results.

\begin{figure}[H]
 \centering
 \includegraphics[width=0.4\textwidth]{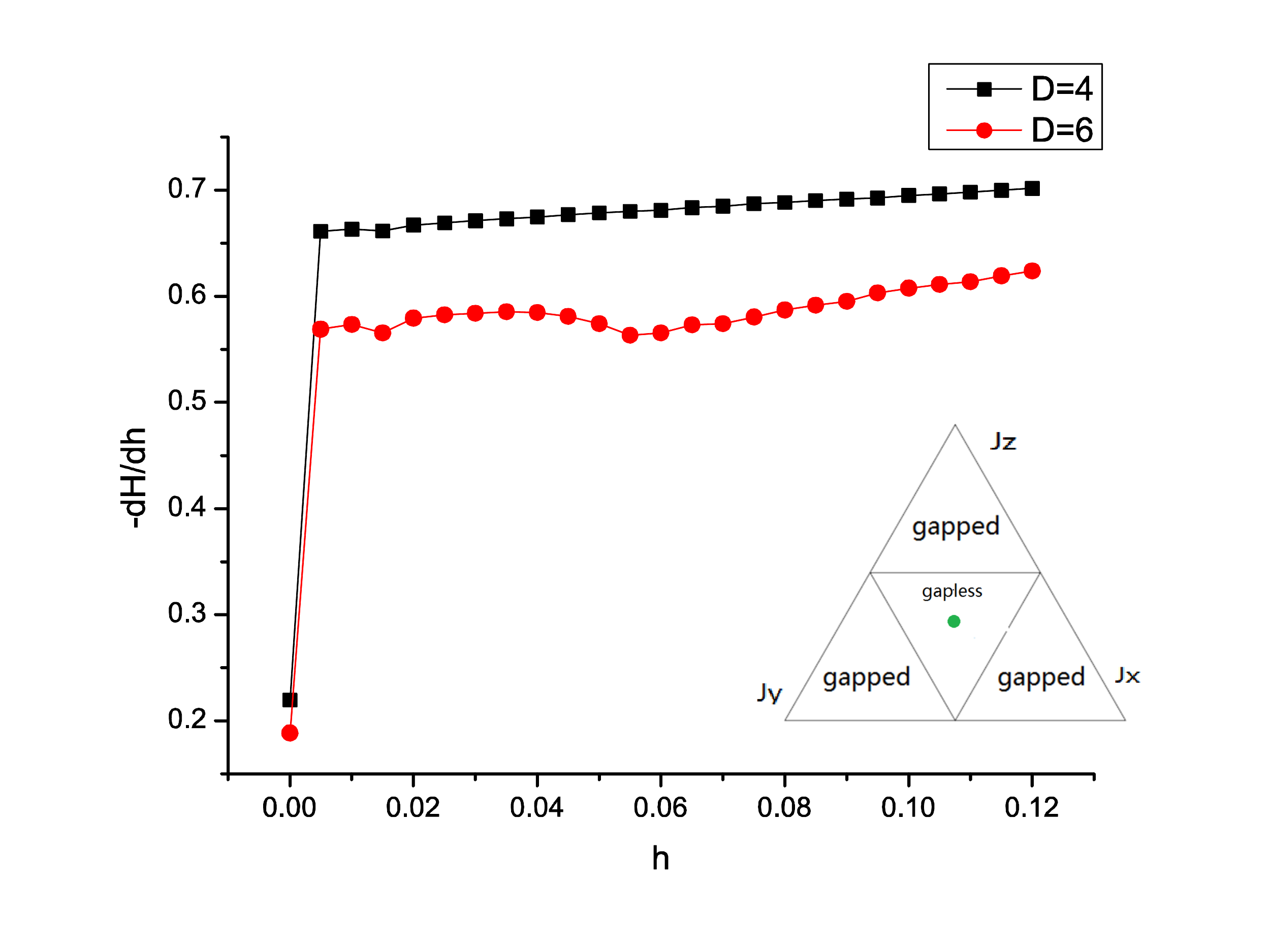}
 \caption{(Color on line) The transition according to the modulation of $-\frac{\partial{H}}{\partial{\vec{h}}}$  with a magnetic field $h$. The Kitaev model is initially in the spin liquid phase with parameters $J_x=Jy=Jz=1/3$, as denoted by the green point. The red dots indicate results for the parameter $D=4$, and the black squares display the results for $D=6$. Although there is numerical difference between the two parameters (without finite scaling for $D$), the transition is sufficiently sharp enough and at the same position. This finding provides strong evidence that there is a first phase transition occurs near $h=0.005$ under finite scaling.}
 \label{fig.3}
\end{figure}

Analytical result\cite{kitaev} suggest that the gapless spin liquid will exhibit a gap upon the addition of a magnetic field in the direction $\textbf{e}_x+\textbf{e}_y+\textbf{e}_z$ ($(1,1,1)$ direction) as a perturbation, which breaks the time-reversal symmetry. Unfortunately, the Kitaev model with magnetic field cannot be solved exactly. We can only study this transition by numerical methods. Our mPEPS calculations show this transition directly by $-\frac{\partial{H}}{\partial{\vec{h}}}$ where $h$ is the magnitude of the magnetic field (see Fig.\ref{fig.3}), and these calculations strongly suggest a first order phase transition occurring at very low magnetic fields, $h < 0.005$.

\textbf{Application to the Kitaev-Heisenberg Model} Having verified the mPEPS method in the Kitaev model, we extend the application of this method to the Kitaev-Heisenberg model. The Hamiltonian of this model can be written as: $H_{HK}=(1-\alpha)H_H+2\alpha H_K$, where $H_H$ is the standard Heisenberg interaction: $\sum_{\gamma=x,y,z}\sigma^{\gamma}\sigma^{\gamma}$ and the $H_K=\sum_{\gamma-link}\sigma^{\gamma}_i\sigma^{\gamma}_j$ is Kitaev interaction. With the Khaliullin transformation\cite{KHmodel}, which performs rotations on different sites, this model can be exactly solved at the point $\alpha=0.5$ and $\alpha=1$. When $\alpha=0.5$, the ground state of the transformed system can be exactly solved, and is found to be a ferromagnetic state. With the inverse Khaliullin transformation, the ferromagnetic state will be transformed to the stripy anti-ferromagnetic state. When $\alpha=1$, this model is equivalent to the Kitaev model, with $J_x=J_y=J_z$, whose ground state can be exactly solved as a gapless spin liquid state. For the parameters beyond $\alpha=0.5$ and $\alpha=1$, the model can not be exactly solved and can only be examined by numerical calculations. Former studies have employed several numerical methods, such as exact diagonalization\cite{KHmodel, Schaffer} and DMRG\cite{Jiang}, and have found three different phases: the Neel order phase, the stripy anti-ferromagnetic phase and the spin liquid phase. These previous calculations, which were based on the systems of relatively small finite size\cite{KHmodel} or quasi one-dimensional\cite{Jiang}, showed that phase transition points occur near $\alpha=0.4$ and $\alpha=0.8$ at zero temperature.

We applied the mPEPS method to this model on an infinite two-dimensional lattice which naturally satisfies the thermodynamic limit. We directly calculate the spin configuration of the ground state with $\alpha=0.5$ which is indeed a ferromagnetic state, as predicted by the exact results. As with the benchmark calculation of the Kitaev model with $J_x=J_y=J_z=1/3$ (corresponding to $\alpha=1$), here we also give $\frac{dH}{d\alpha}$, which can be used to determine the phase transition point, as shown in Fig.4.

In our infinite system with six-site periodicity, the second transition point is dramatically different from the former result($\alpha=0.8$), while the first transition point, at $\alpha=0.46$, is slightly different from $\alpha=0.4$. To confirm our calculation, we also determine the NN correlation $\langle \vec{\sigma}\vec{\sigma}\rangle$, which is used as an indication of the phase transition in reference\cite{KHmodel}. The results also support the occurrence of a phase transition at the same position. In our calculations, the second phase transition point occurs at $\alpha\approx0.98$ for $D=4$, and at $\alpha\approx0.99$ for $D=6$. This findings suggest that the transition point will tend to $\alpha=1$ with the finite scaling for $D$. Our results indicate that the gapless spin liquid phase will be destroyed by the Heisenberg perturbation very quickly, and thus this phase is not robust for this type (Heisenberg) of perturbation. The direct implication of our results indicates that it would be very difficult to observe a Kitaev-type gapless spin liquid in iridate Na$_2$IrO$_3$.
The findings of a decreasing spin liquid phase in the infinite system (at the thermodynamics limit), compared to the finite exact diagonalization and DMRG in quasi one-dimension with cylinder boundary condition, is not so surprising. The size of the system plays a subtle role in the spin liquid phase. Z. Meng \emph{et al} have reported\cite{ZYMeng} a spin liquid phase between the semimetal phase and an antiferromagnetic Mott insulator in the Fermi-Hubbard model on the honeycomb lattice, with the inclusion of 648 sites, by using projective QMC method. However, when S. Sorella \emph{et al}\cite{japanese} reconsidered this problem by including up to 2592 sites, the evidence of spin liquid was lost. In our case, the infinite lattice may play a similar role.

\begin{figure}[H]
 \centering
 \includegraphics[width=0.5\textwidth]{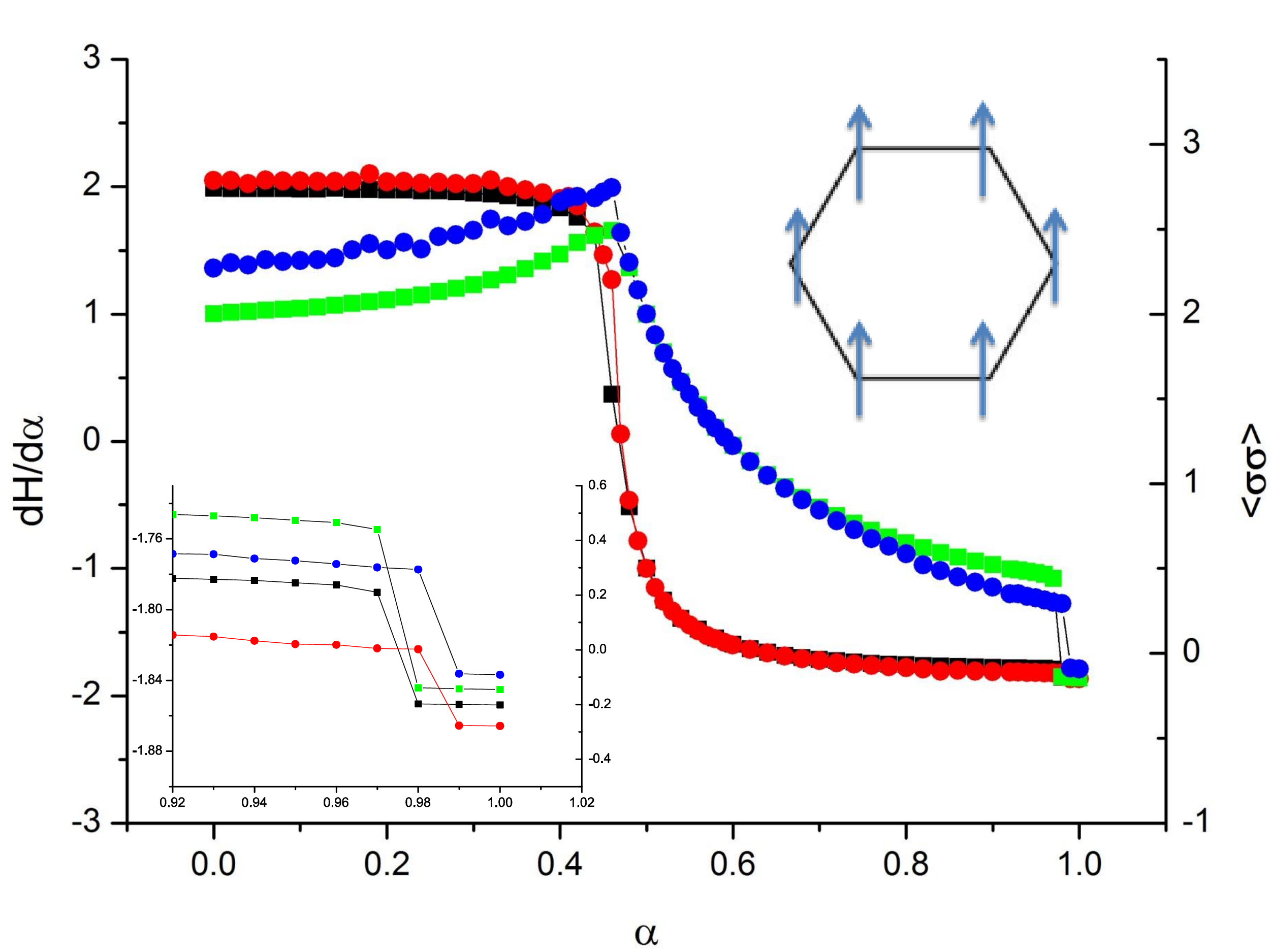}
 \caption{(Color online) The values of $\frac{dH}{d\alpha}$ (a) and the NN correlation $\langle \vec{\sigma}\vec{\sigma}\rangle$ (b) versus $\alpha$ for $D=4$ (black and green squares) and $D=6$ (red and blue dots). The left lower inset shows the magnification near $\alpha=1$ and the right upper inset displays the configuration of the spin under the Khaliullin transformation for $\alpha=0.5$.  It is clear that two phase transitions occur. The first transition point is near $\alpha=0.46$ which is only slightly different with $0.4$. However, the second phase transition occurs at $\alpha\approx0.98$ which is qualitatively different from $0.8$. Our results show that the spin liquid region is very narrow and that the gapless spin liquid will be rapidly destroyed by the Heisenberg perturbation.}
 \label{fig.4}
\end{figure}

To summarize, we have introduced a novel type of TN (mPEPS) algorithm to attack the highly frustrated spin systems. Our method is successful in simulating the Kitaev model, with respect to both the energy and the correlations of the ground state. Having established this method, we applied it to the Kitaev-Heisenberg model. Our calculations show that the spin liquid region is very narrow, and thus this state would be very difficult to observe experimentally. The success of our method opens the possibility of considering virtual particles in a tensor network with more complex statistical characteristics, such as the fractional statistics of anyons. These new statistics may be convenient for studying the system which excitations have the similar statistics.

We acknowledge support from the Chinese National Fundamental Research Program 2011CB921200 and the Fundamental Research Funds for Central Universities Nos. WK2470000006, WK2470000004, WJ2470000007 and NSFC11105135.

\clearpage
\setcounter{figure}{0}
\renewcommand\thefigure{A\arabic{figure}}
\setcounter{equation}{0}
\renewcommand\theequation{A\arabic{equation}}
\subsection{\textbf{Supplementary}}

\subsubsection{\textbf{Imaginary Time Evolution}}

After chosen the variational space of the state, i.e., mPEPS with given parameter $D$ and $d$, we use the imaginary time evolution method\cite{PEPS1,PEPS2,PEPS3} to find the ground state. The imaginary time evolution could be summarized as the following formula:

\begin{equation*}
   |\psi_{g}\rangle=\lim_{t\rightarrow\infty}\frac{e^{-Ht}|\psi_0\rangle}{\langle\psi_0|e^{-Ht}e^{-Ht}|\psi_0\rangle}
\end{equation*}

The $|\psi_0\rangle$ is the initial state, which can be chosen randomly, after long enough imaginary time evolution, the final normalized state will be close to the ground state with well controlled precision, as long as the initial state is not orthogonal with the ground state (with zero measurement to chosen these states). Generally, we can only deal with the local evolution operators (the time evolution of a system is a global operator), and the Trotter expansion can replace the global operators by local operators with controllable precision\cite{PEPS1,PEPS2,PEPS3}. For example, in Kitaev model, Hamiltonian $H$ could be split as $H=H_{x}+H_{y}+H_{z}$, where $H_{i},(i=x,y,z)$ is the part of Hamiltonian acting on $i$-bond. With the help of Trotter expansion, we could split the evolution operator in second order as:
\begin{eqnarray*}
   && e^{-Hdt}=e^{-(H_{x}+H_{y}+H_{z})dt}\\
   && =e^{-H_{x}dt}e^{-H_{y}dt}e^{-2H_{z}dt}e^{-H_{y}dt}e^{-H_{x}dt}+O(dt^{3})
\end{eqnarray*}

\begin{figure}[H]
 \centering
 \includegraphics[width=0.5\textwidth]{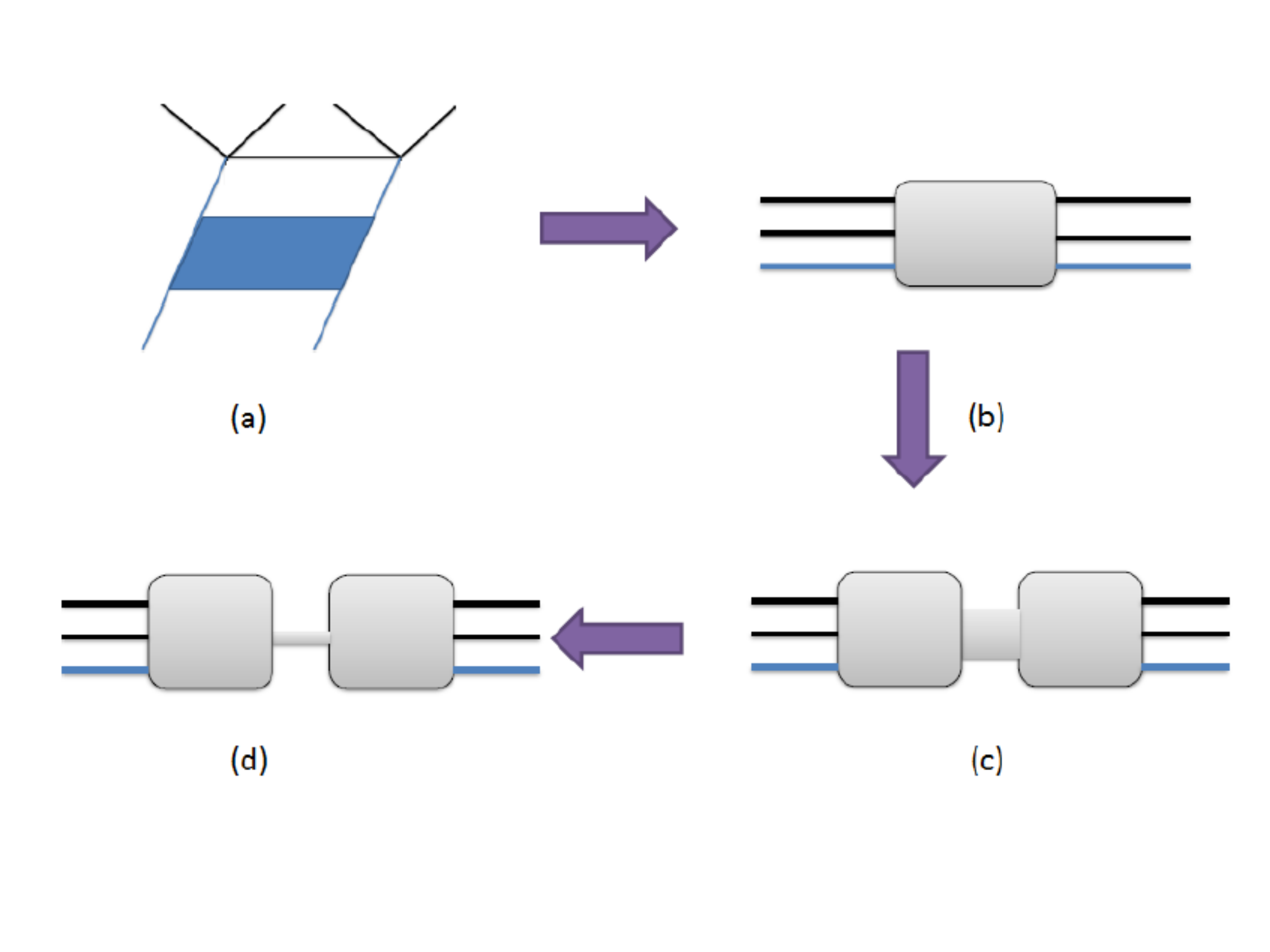}\\
 \caption{SVD in imaginary time evolution: (a) Act the evolution operator $e^{-dtH}$ (the parallelogram blue tensor) on two contracted tensors. Then we obtain a bigger tensor as indicated by (b). (c) Use SVD to cut the big tensor and (d) reserve only D biggest singular values to make the tensors back to their original dimension. For convenience, before any contract processes of the tensors, we contract the diagonized matrix, which is the approximation of the environment, into the nearest tensors.}
 \label{fig:SVDiTE}
\end{figure}

With the Trotter expansion, it seems that we just need to act the local evolution operators on the initial state with the given order and obtain the ground state finally. However, the dimension of the bond $D$ will increase exponentially with the evolution time $t$, if local operators are applied without approximation. Therefore, as in the general TN algorithm, approximations should be made after each step of evolution. One of the simplest way to cut the tensors back to the initial parameter space is the singular value decomposition (SVD), which maintain the important part of the tensor (see Fig.\ref{fig:SVDiTE}). It is straightforward to use SVD in the case of PEPS, while it needs extra operations in the case of mPEPS (also in fPEPS) with fermionic statistics. Since the tensors in mPEPS are set with even parity, the matrix in the SVD process can be re-indexed as block diagonal matrix according to parity of the tensor's indices (one block with odd parity and the other with even parity), then we can apply the SVD and cut off in each of the block.

\subsubsection{\textbf{Contract}}

After the imaginary time evolution, we suppose that we have found the ground state. Then the key task is to find the values of some physical quantities which can be compared with the experiment. This task is realized by the contract processes, which including the contract of tensors and entangled pairs. The physical quantities in a given ground state $|\psi_{g}\rangle$ can be expressed as

\begin{equation*}
   \langle O\rangle=\frac{\langle \psi_{g}|\hat{O}|\psi_{g}\rangle}{\langle \psi_{g}|\psi_{g}\rangle}
\end{equation*}

For simplicity, we use $\langle \psi_{g}|\psi_{g}\rangle$ as an example to show how to contract to calculate the physical quantities. The calculation of physical quantities is similar. Suppose we have a 2D honeycomb lattice and each site is labeled with two indices i,j. We have a state in the form of projected entangled-pair State:
\begin{equation*}
    |\psi\rangle=\prod_{ij}P^{[ij]}\prod_{i_1j_1i_2j_2}\mbox{EPR}_{[i_1j_1][i_2j_2]} \, ,
\end{equation*}
where $\mbox{EPR}_{[i_1j_1][i_2j_2]}=\sum_{v}|v\rangle_{[i_1j_1]}|v\rangle_{[i_2j_2]}$ and $[i_1j_1][i_2j_2]$ denotes the edge e, linking site $[i_1j_1]$ and site $[i_2j_2]$.  Normally, translational invariance of tensors will be imposed in infinite lattice. Actually, we have the corresponding bra vector as
\begin{equation*}
    \langle \psi|=\prod_{i_1j_1i_2j_2}{\mbox{EPR}_{[i_1j_1][i_2j_2]}^{\dag}}\prod_{ij}{P^{[ij]{\dag}}} \, ,
\end{equation*}
where $\mbox{EPR}^{\dag}_{[i_1j_1][i_2j_2]}=\sum_{v}\langle v|_{i_2j_2}\langle v|_{i_1j_1}$. The product order is reversed which will make a difference in the contract procedure below. The scalar product of bra vector and ket vector is:
\begin{eqnarray*}
&&  \langle \psi|\psi\rangle=\prod_{i_1j_1i_2j_2}{\mbox{EPR}_{[i_1j_1][i_2j_2]}^{\dag}}\prod_{ij}{P^{[ij]{\dag}}}\\
&&  \prod_{ij}{P^{[ij]}}\prod_{i_1j_1i_2j_2}{\mbox{EPR}_{[i_1j_1][i_2j_2]}}
\end{eqnarray*}

Because we are considering mPEPS where tensors have even parity, it is free to change the orders. We would like to rearrange the order as:
\begin{eqnarray*}
&&  \langle \psi|\psi\rangle=\prod_{i_1j_1i_2j_2}{\mbox{EPR}_{[i_1j_1][i_2j_2]}^{\dag}}\\
&&  \prod_{ij}{P^{[ij]{\dag}}P^{[ij]}}\prod_{i_1j_1i_2j_2}{\mbox{EPR}_{[i_1j_1][i_2j_2]}}
\end{eqnarray*}

The product of $P^{[ij]\dag}P^{[ij]}$ will contract over physical indices and we symbolize it as $K_{ij}$.
\begin{small}
\begin{eqnarray*}
&&  K_{ij}=P^{[ij]\dag}P^{[ij]}\\
&& =\sum T^{[ij]\ast}_{\overline{p}\overline{v_x}\overline{v_y}\overline{v_z}}|\overline{v_z}\overline{v_y}\overline{v_x}\rangle\langle \overline{p}|
^{[ij]}_{pv_xv_yv_z}|p\rangle\langle v_xv_yv_z|\\
&&  =\sum T^{[ij]\ast}_{p\overline{v_x}\overline{v_y}\overline{v_z}}T^{[ij]}_{pv_xv_yv_z}|\overline{v_z}\overline{v_y}\overline{v_x}\rangle\langle v_xv_yv_z|\\
&&  \rightarrow \sum K^{[ij]}_{\overline{v_x}v_x\overline{v_y}v_y\overline{v_z}v_z}|\overline{v_x}\rangle\langle v_x|\overline{v_y}\rangle\langle v_y|\overline{v_z}\rangle\langle v_z| \, ,
\end{eqnarray*}
\end{small}
where repeated indices mean contraction. Notice that $K^{[ij]}_{\overline{v_x}v_x\overline{v_y}v_y\overline{v_z}v_z}$ is not just the contraction of $T^{[ij]\ast}_{p\overline{v_x}\overline{v_y}\overline{v_z}}$ and $T^{[ij]}_{pv_xv_yv_z}$, but with extra minus signs resulting from commutation of operators.

Therefore, the scalar product is
\begin{equation*}
   \langle \psi|\psi\rangle=\prod_{i_1j_1i_2j_2}{\mbox{EPR}^{\dag}_{[i_1j_1][i_2j_2]}}\prod_{ij}{K_{ij}}\prod_{i_1j_1i_2j_2}{\mbox{EPR}_{[i_1j_1][i_2j_2]}}
\end{equation*}

\begin{figure}[H]
 \centering
 \includegraphics[width=0.5\textwidth]{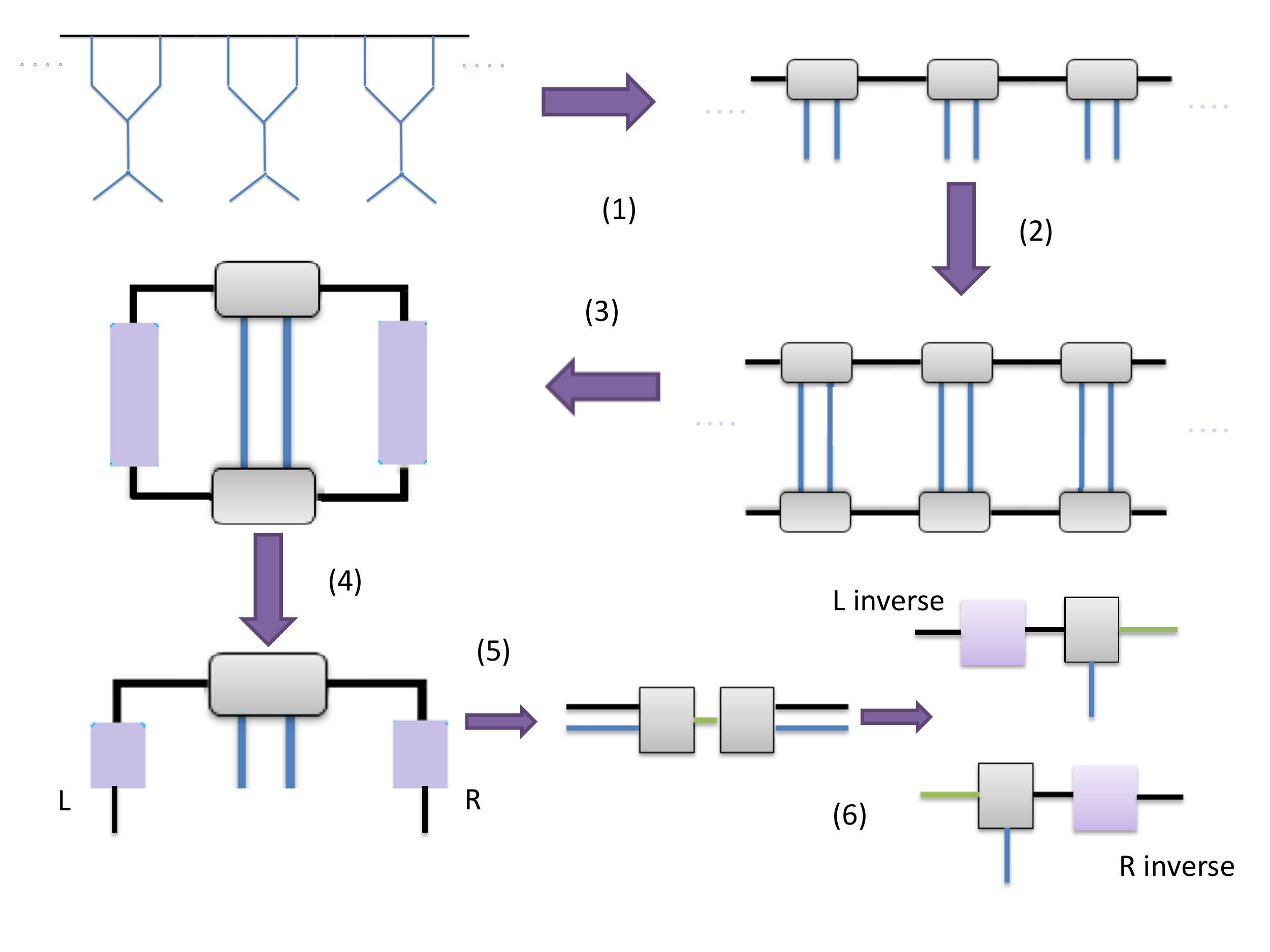}\\
 \caption{contract procedures: (1) Contract boundary tensors with one row of K tensors; (2) trace the new row of tensors with their conjugate; (3) Seek left and right environment tensors by iterative product; (4) Obtain L and R tensor by cutting left and right environment using eigenvalue decomposition, Contract L, the fixed tensor, and R together; (5) Use SVD to make the approximation; (6) Contract L's inverse and R's inverse tensors respectively.}
 \label{fig:contract}
\end{figure}

What we should do next is to contract over all virtual indices, ie. to product all the $K$ tensors by \mbox{EPR}s. Even though we consider here is a infinite lattice, we suppose there are boundaries in very remote areas. We contract row by row many times from boundaries until the boundary tensors are stable. After we contract the boundary with a row of tensors, the boundary tensors will get thicker, like the case of imaginary time evolution, and we need to do approximations to cut them back to their initial shape. Normally we use a specific form of boundary tensors to make the approximation, for the sake of simplicity (see Fig.\ref{fig:contract} for details).

\begin{figure}[H]
 \centering
 \includegraphics[width=0.5\textwidth]{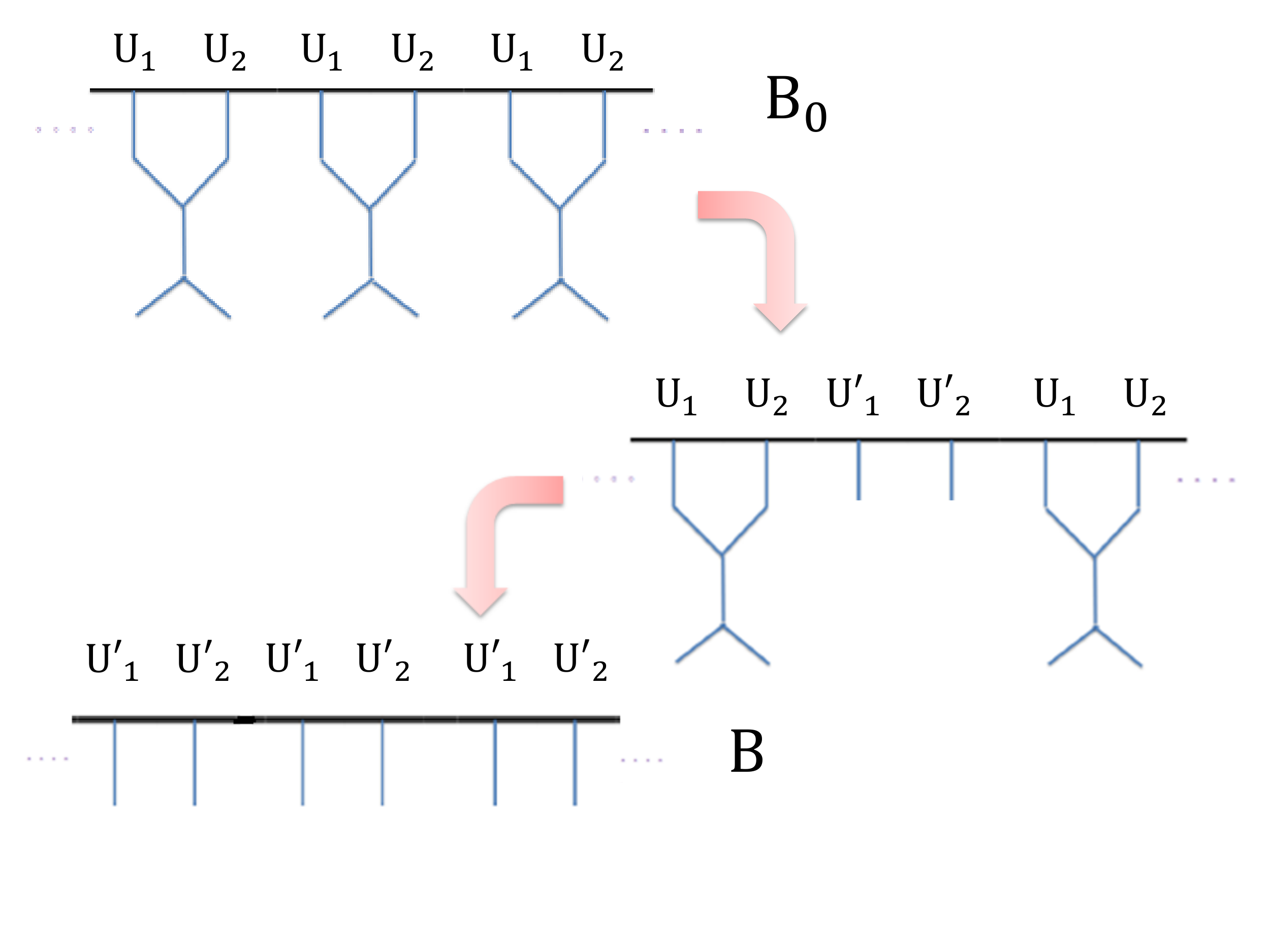}    \\
 \caption{The scheme of approximation in the language of variational method. Use the second row of boundary tensors to approximate the first row. Substitute the first row with the third row. And keep on this procedures until stable.}
 \label{fig:Approxi}
\end{figure}

The main tool used to make this approximation is still SVD: We fix two tensors, and trace over left and right parts of boundary tensors to gain the left and right environments of the two fixed tensors. Now we contract over left environment tensor, fixed tensors, right environment tensors together to become a bigger tensor. We use SVD to cut the bigger tensor, then cancel the environment effect by contracting inverse tensors of the left and right environments. It is extremely obvious if you write down the approximation judgement explicitly:
\begin{small}
\begin{eqnarray*}
&&  \min_{B}(\mbox{Tr}(B-B_{0})^{\dag}(B-B_{0}))   \\
&&  =\min_{U^{\prime}_1U^{\prime}_2}\mbox{Tr}((G_LU^{\prime}_1U^{\prime}_2G_R-G_LGG_R)^{\dag}(G_LU^{\prime}_1U^{\prime}_2G_R-G_LGG_R)) \\
&&  =\min_{U^{\prime}_1U^{\prime}_2}(\mbox{Tr}(G_R^{\dag}U^{\prime\dag}_2U^{\prime\dag}_1G_L^{\dag}G_LU^{\prime}_1U^{\prime}_2G_R) \\
&&  -\mbox{Tr}(G_R^{\dag}U^{\prime\dag}_2U^{\prime\dag}_1G_L^{\dag}G_LGG_R)
\mbox{Tr}(G_R^{\dag}G^{\dag}G_L^{\dag}G_LU^{\prime}_1U^{\prime}_2G_R)+    \\
&&  \mbox{Tr}(G_R^{\dag}G^{\dag}G_L^{\dag}G_LGG_R))    \\
&&  =\min_{U^{\prime}_1U^{\prime}_2}(\mbox{Tr}(R^{\dag}U^{\prime\dag}_2U^{\prime\dag}_1LU^{\prime}_1U^{\prime}_2)
\mbox{Tr}(R^{\dag}U^{\prime\dag}_2U^{\prime\dag}_1LG) \\
&&  -\mbox{Tr}(R^{\dag}G^{\dag}LU^{\prime}_1U^{\prime}_2)+\mbox{Tr}(R^{\dag}G^{\dag}LG)) \, ,
\end{eqnarray*}
\end{small}
where $B_{0}$ is the old boundary contracting with a row of tensors, B is the new boundary with only two tensors different from the old boundary. G is the unit cell of $B_0$. $G_L$ $G_R$ are perspectively the left and right part of G in the row of $B_0$. L is just $G_L^{\dag}G_L$ and R is just $G_R^{\dag}G_R$. The above equation is just binary form of $U_1^{\prime}$ and $U_2^{\prime}$. Obviously, $\mbox{Tr}(B-B_{0})^{\dag}(B-B_{0})$ reaches its minimum when $\sqrt{R^{\dag}}U_1^{\prime}U_2^{\prime}\sqrt{L}$ best approximates $\sqrt{R^{\dag}}G\sqrt{L}$. See also Fig.\ref{fig:Approxi}.

This example of $\langle \psi|\psi\rangle$ showed our method for contract. It is similar to calculate average value of local operators $\langle \psi|\hat{O}|\psi\rangle$, except that the sites where operators lie should be kept until final contract.

There is one more point worth mentioning: When we use mPEPS, we choose the virtual annihilation and creation operators to be anti-commutative. So we must be extremely careful about the minus sign when we contract over virtual indices. Similarly and more easily to overlook, in SVD, minus signs also appear when we change the order of virtual indices in order to obtain the original shape of tensors. It could be explained as equations below. We take a simplified version as an example. T is a big tensor which needs to be split by SVD.

\begin{small}
\begin{eqnarray*}
&&  T_{\bar{u}u,\bar{v}v} \\
&&  = \sum_{\bar{s}s}U^{\prime}_{\bar{u}u,\bar{s}s}S_{\bar{s}s}V^{\prime}_{\bar{s}s,\bar{v}v} \\
&&  = \sum_{\bar{s}s}U_{\bar{u}u,\bar{s}s}V_{\bar{s}s,\bar{v}v}  \\
&&  =\sum_{\overline{s1}s1\overline{s2}s2}U_{\bar{u}u,\overline{s1}s1}\delta_{\overline{s1}s1,\overline{s2}s2}V_{\overline{s2}s2,\bar{v}v}  \\
&&  =\overline{\mbox{EPR}}^{\dag}(\sum_{\overline{s1}s1\overline{s2}s2}U_{\bar{u}u,\overline{s1}s1}|\overline{s2}\overline{s1}\rangle\langle s1s2|
_{\overline{s2}s2,\bar{v}v})\mbox{EPR}    \\
&&  =\overline{\mbox{EPR}}^{\dag}(\sum_{\overline{s1}s1\overline{s2}s2}U_{\bar{u}u,\overline{s1}s1}(-)^{(p(\overline{s1})+p(s1))p(\overline{s2})}  \\
&&  |\overline{s1}\rangle\langle s1||\overline{s2}\rangle\langle s2|V_{\overline{s2}s2,\bar{v}v})\mbox{EPR}  \\
&&  =\overline{\mbox{EPR}}^{\dag}(\sum_{\overline{s1}s1\overline{s2}s2}U_{\bar{u}u,\overline{s1}s1}(-)^{(p(\overline{s1})+p(s1))p(\overline{s1})}  \\
&&  |\overline{s1}\rangle\langle s1||\overline{s2}\rangle\langle s2|V_{\overline{s2}s2,\bar{v}v})\mbox{EPR}  \\
&&  =\overline{\mbox{EPR}}^{\dag}(\sum_{\overline{s1}s1\overline{s2}s2}U^{\prime\prime}_{\bar{u}u,\overline{s1}s1} \\
&&  |\overline{s1}\rangle\langle s1||\overline{s2}\rangle\langle s2|V_{\overline{s2}s2,\bar{v}v})\mbox{EPR}  \\
&&  =contract\ U^{\prime\prime}\ tensor\ with\ V\ tensor
\end{eqnarray*}
\end{small}

The first equation is normal SVD; the second is to absorb singular values; and rest equations are to extract \mbox{EPR} from tensors. \mbox{EPR} is $\sum_{s2=s1}|s2\rangle|s1\rangle$, correspondingly $\mbox{EPR}^{\dag}=\sum_{s2=s1}\langle s1|\langle s2|$, and so as their Hermitian conjugate. In the equations, you can see that a minus sign $(-)^{(p(\bar{s1})+p(s1))p(\bar{s1})}$ show up. Note that $p(\bar{s2})$ in the fifth equation is replaced by $p(\bar{s1})$ since it is not zero only when $s1=s2$ and $\bar{s1}=\bar{s2}$. p(*) is the function that maps quantum numbers to parity. U and V is the tensors that $U^{\prime}$ and $V^{\prime}$ absorb square of the singular values respectively. $U^{\prime\prime}$ is the tensor after U absorbs $(-)^{(p(\bar{s1})+p(s1))p(\bar{s1})}$.

\subsubsection{\textbf{Additional Calculation of the Kitaev Model}}
In order to shows the stability of our results, we increase the dimension of the truncation n the contract from $D_{cut}=3^2$ to $D_{cut}=7^2$. We can see in the following figure that when $D_{cut} > 4^2$ with $D=8$, the NN correlation of the system is relatively stable.

\begin{figure}[H]
 \centering
 \includegraphics[width=0.4\textwidth]{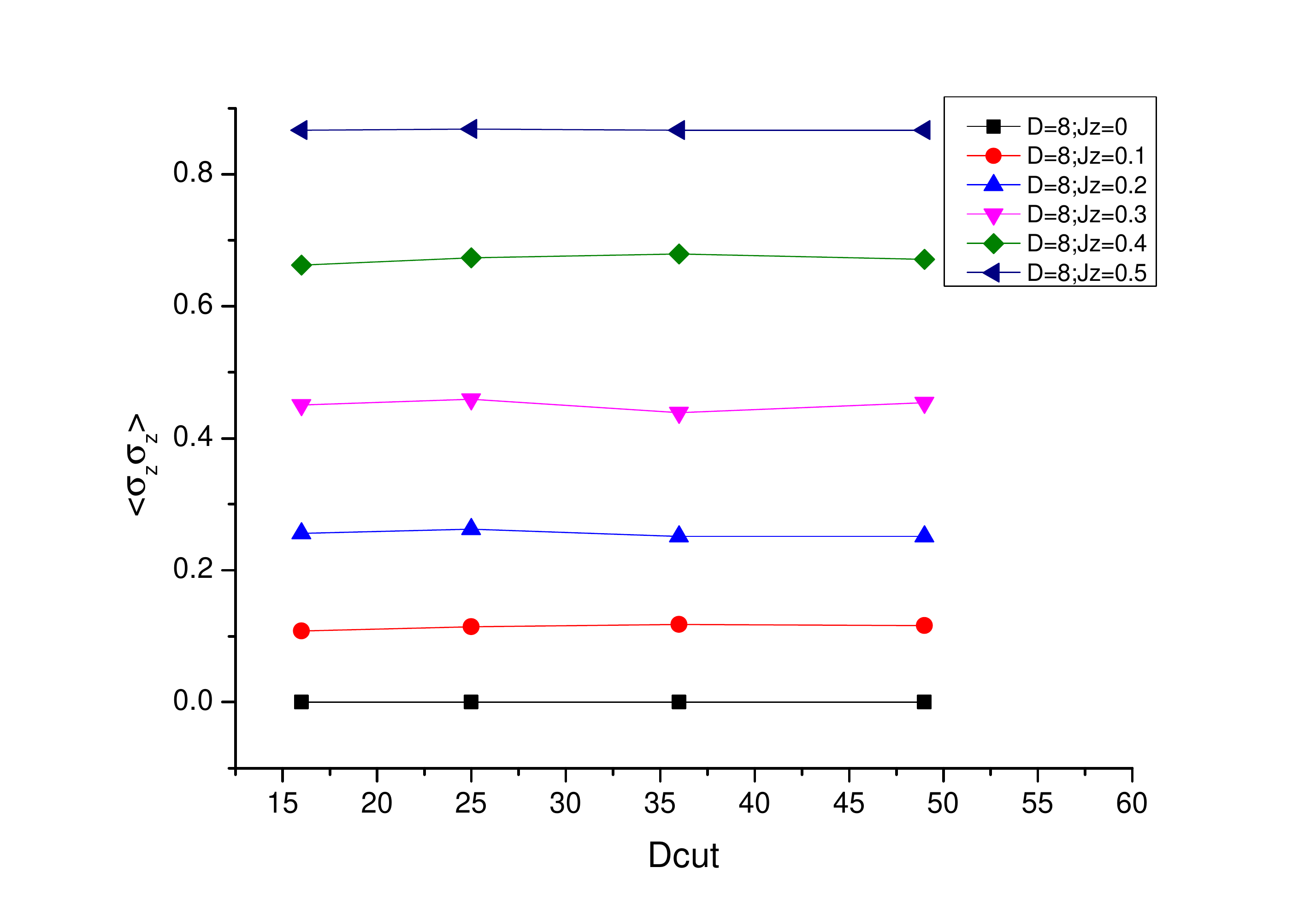}
 \caption{The diagram shows that when $D=8$, the NN correlations $\langle \sigma^z\sigma^z\rangle$ are stable as $D_{cut}$ increases from $D_{cut}=4^2$ to $D_{cut}=7^2$.}
 \label{fig:stable}
\end{figure}

With the stable algorithm, we calculate the energy of the Kitaev model which is well agree with the exact energy
\begin{figure}[H]
 \centering
 \includegraphics[width=0.4\textwidth]{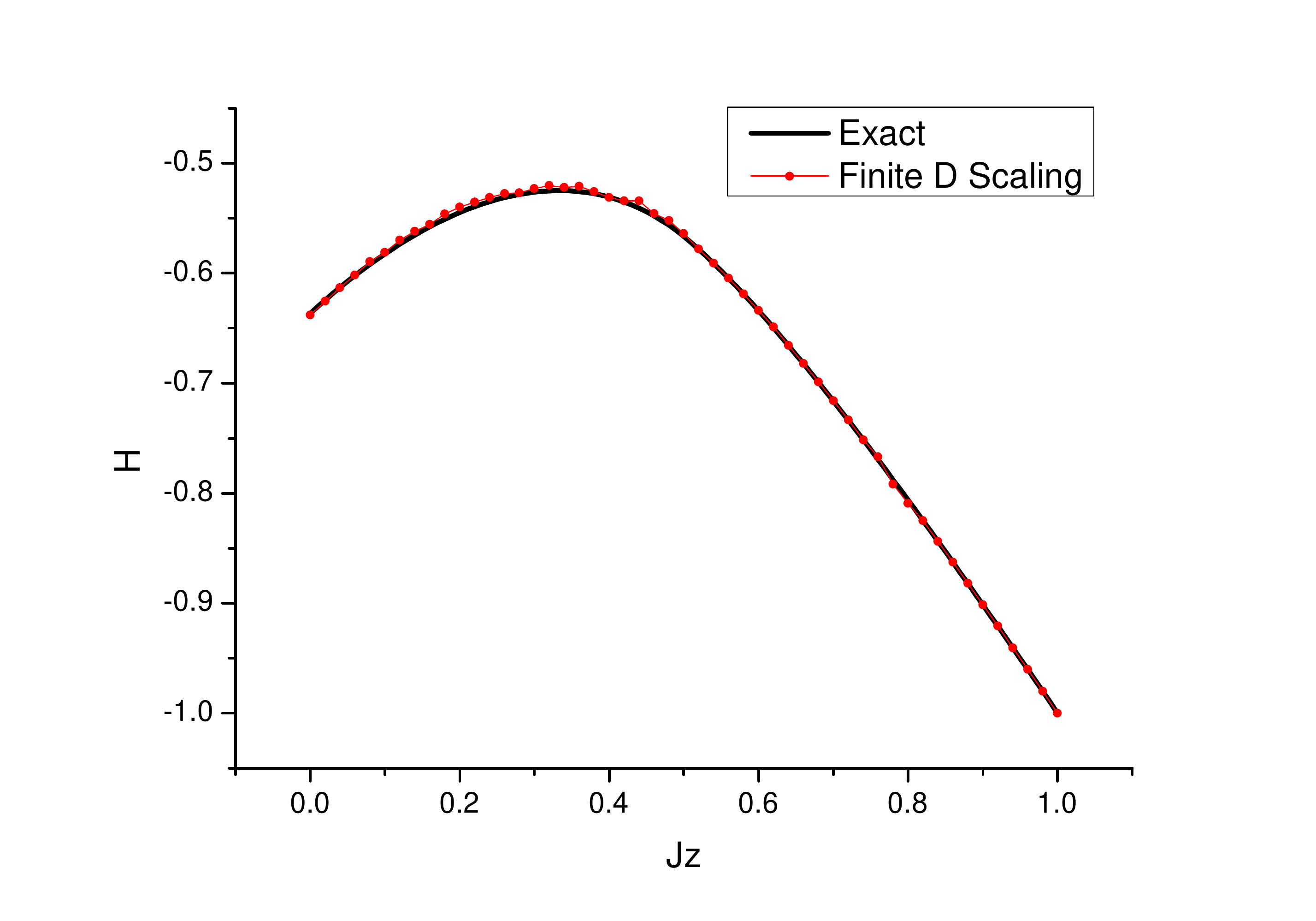}
 \caption{The diagram shows the ground state energy of Kitaev model by fPEPS.(ST stands for exact results. FS stands for the results after finite scaling of D.)}
 \label{fig:KitaevFSH}
\end{figure}

Since the limit of our calculation, we should do some finite scaling. However, the finite scaling of D is not clear and well defined. Here we use linear of polynomial fit with $1/D$, as showed below.
\begin{figure}[h]
  \centering
  \includegraphics[width=0.4\textwidth]{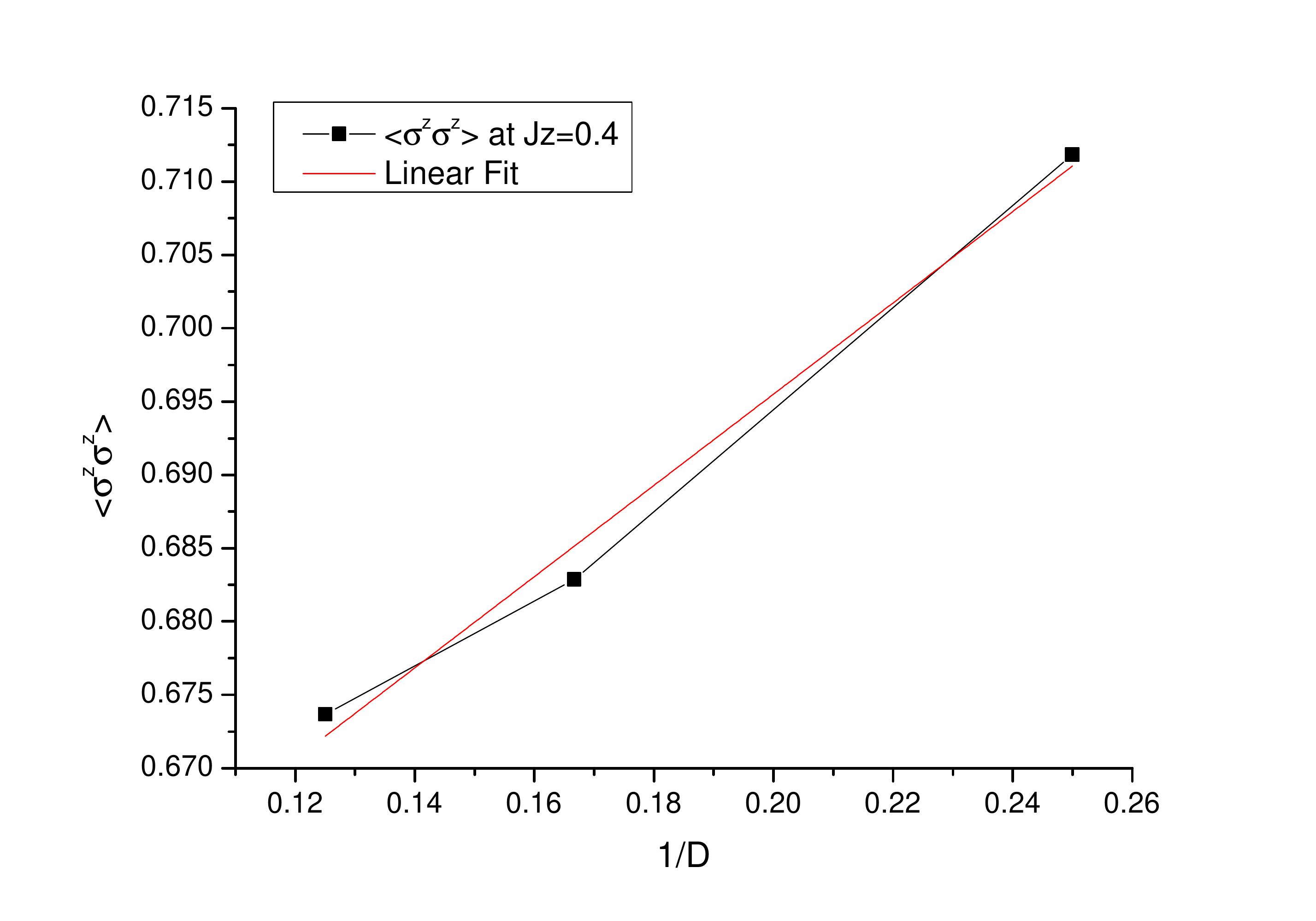}\\
  \caption{Finite D Scaling of $\langle \sigma^z\sigma^z\rangle_z$ at Jz=0.4. We choose it to linear-fit with $1/D$.}
\end{figure}

\begin{figure}[H]
 \centering
 \includegraphics[width=0.3\textwidth]{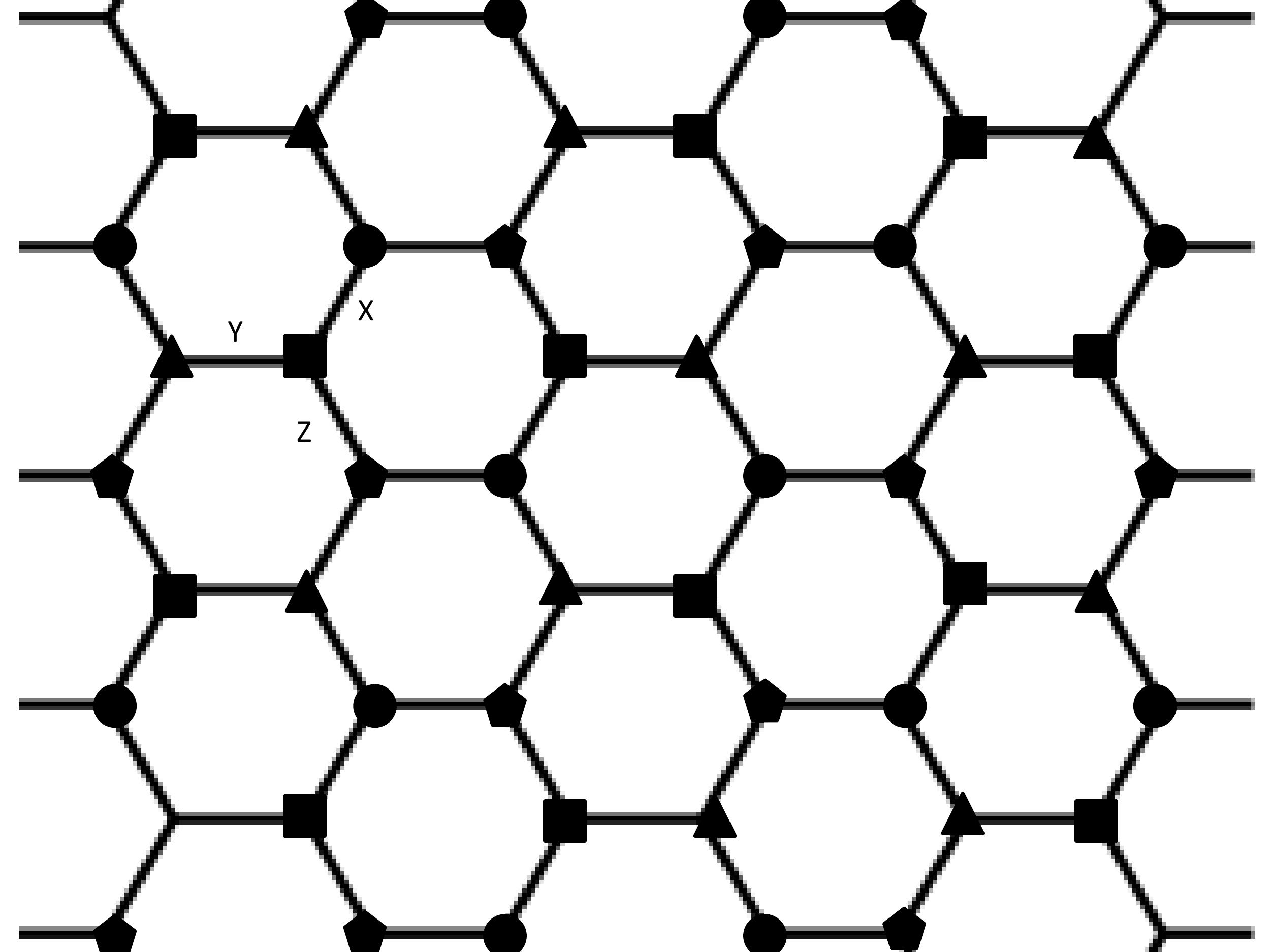}
 \caption{Khaliullin Transformation: The spins on square sites are fixed. The spins on circle sites are rotated by $\pi$ along x bond. The spins on triangle sites are rotated by $\pi$ along y bond. The spins on pentagon sites are rotated by $\pi$ along z bond.}
\end{figure}

\subsubsection{ \textbf{Khaliullin Transformation for Kitaev-Heisenberg Model}}

This transformation first proposed by Khaliullin and Chaloupka not only makes the anti-ferromagnetic state to ferromagnetic state, but also makes Kitaev-Heisenberg model exactly solvable at $\alpha=1/2$. This transformation requires different spins in the lattice to rotate about different axis. To be specific, we first choose a set of spins which are positioned on third nearest neighbor sites at opposite corners of the hexagons throughout the lattice, and hold these spins fixed. We next rotate the three nearest spins by $\pi$ about the spin axis corresponding to the bond which connects it to the fixed spin. The effect of this operation therefore transforms the Heisenberg Hamiltonian to a mixture of Heisenberg Hamiltonian and Kitaev Model, and Kitaev Hamiltonian invariant.

\begin{eqnarray*}
&&  H_{H}\rightarrow-H_{H}+2H_{K}\\
&&  H_{K}\rightarrow H_{K}
\end{eqnarray*}

Thus the Kitaev-Heisenberg model is transformed as:

\begin{equation}\label{eqn:oldHKmodel}
   H\rightarrow-(1-\alpha)H_{H}-4(\alpha-1/2)H_{K}
\end{equation}

It is obvious to find out in the transformed Hamiltonian, when $\alpha=1/2$, it is just $H=-1/2H_{H}$ Heisenberg Hamiltonian with a negative coupling coefficient. So the state at $\alpha=1/2$ after the transformation is a ferromagnetic state. We can operate an inverse transformation to send the ferromagnetic state back to an anti-ferromagnetic state. As we have explained before, the anti-ferromagnetic state is harder to calculate, since stable results are not easily required. So in the letter, we just use the Hamiltonian after the transformation to discuss the quantum phase diagram of Kitaev-Heisenberg model.

In order to weaken the influence by Heisenberg interaction and make the spin liquid region broader and easier to see, we re-parameterize the Kitaev-Heisenberg model to:
\begin{equation}\label{eqn:newHKmodel}
    H\rightarrow-\frac{1-\alpha^\prime}{3}H_{H}-4(\alpha^\prime-1/2)H_{K}
\end{equation}

The new parameter $\alpha^\prime$ and the old one $\alpha$ can be related by a simple formula, if $H_K$ is proportional to $H_H$ to the same extent. Besides, the new Hamiltonian Eq.\ref{eqn:newHKmodel} preserves the same behaviors as Eq.\ref{eqn:oldHKmodel} at $\alpha^\prime=1$ and $\alpha^\prime=1/2$. In the paper, we discuss Eq.\ref{eqn:newHKmodel}, and note that the spin liquid region of Eq.\ref{eqn:oldHKmodel} is actually much much narrower. The $\alpha^\prime$ region of spin liquids is $0.98< \alpha^\prime < 1$, as the part below suggests. Correspondingly, the real parameter region of spin liquids before re-parameterization are around $0.997\langle \alpha\langle 1$. In the paper, we just use the new Hamiltonian Eq.\ref{eqn:newHKmodel} to further our discussion.

\begin{figure}[H]
 \centering
 \includegraphics[width=0.3\textwidth]{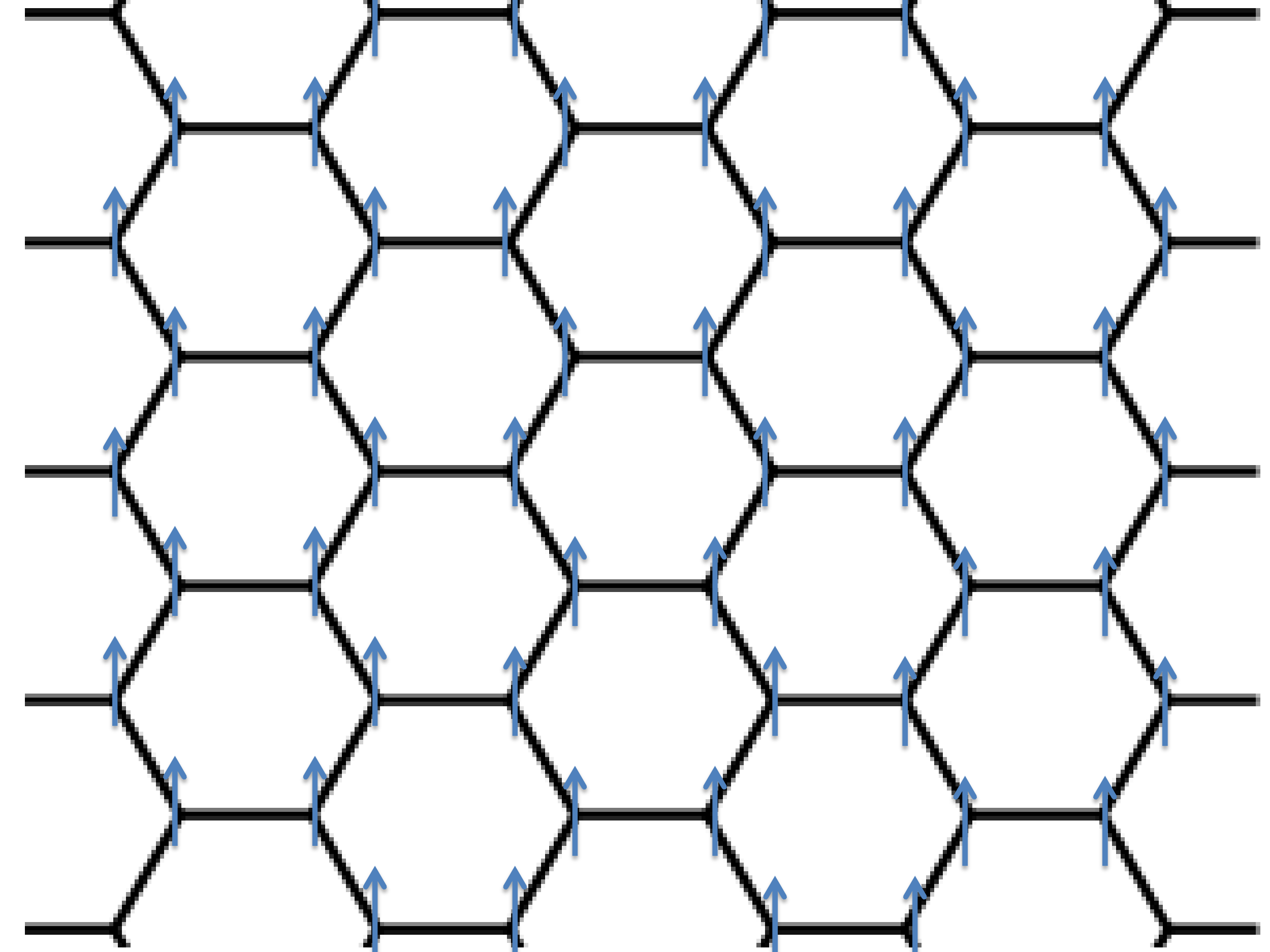}
 \caption{The above figure shows the spin orientation of $\alpha=1/2$ Kitaev-Heisenberg model after Kahliullin transformation. And with the inverse transformation, the spin pattern returns to stripy anti-ferromagnetic order.}
 \label{fig:orientation}
\end{figure}

The configuration of the ground state of the Kitaev-Heisenberg model with $\alpha=0.5$ can be calculated by our PEPS method as Fig.\ref{fig:orientation}

\end{document}